\documentclass[prb,preprint,showpacs,showkeys,amsmath,amssymb,superscriptaddress]{revtex4}

\usepackage{graphicx}
\usepackage{dcolumn}
\usepackage{bm}
\usepackage{graphicx}
\usepackage{epsfig}
\usepackage{amsmath}
\usepackage{natbib}

\begin{document}

\draft
\title{Interplay between magnetism and energetics in FeCr alloys from a predictive non-collinear magnetic tight-binding model}

\author{R. Soulairol}
\affiliation{DEN-Service de Recherches de M\'etallurgie Physique, CEA, Universit\'e Paris-Saclay, F-91191 Gif-sur-Yvette, France}

\author{C. Barreteau}
\affiliation{SPEC, CEA, CNRS, Universit\'e Paris-Saclay, CEA Saclay 91191 Gif sur Yvette, France}
\affiliation{DTU NANOTECH, Technical University of Denmark, {\O}rsteds Plads 344, DK-2800 Kgs. Lyngby, Denmark}

\author{Chu-Chun Fu}
\affiliation{DEN-Service de Recherches de M\'etallurgie Physique, CEA, Universit\'e Paris-Saclay, F-91191 Gif-sur-Yvette, France}

\date{\today}

\begin{abstract}
Magnetism is a key driving force controlling several thermodynamic and kinetic properties of Fe-Cr systems. We present a newly-developed TB model for Fe-Cr, where magnetism is treated beyond the usual collinear approcimation. A major advantage of this model consists in a rather simple fitting procedure. In particular, no specific properties of the binary system is explicitly required in the fitting database. 
The present model is proved to be accurate and highly transferable for electronic, magnetic and energetic properties of a large variety of structural and chemical environments: surfaces, interfaces, embedded clusters, and the whole compositional range of the binary alloy. The occurence of non-collinear magnetic configurations caused by magnetic frustrations is successfully predicted. 
The present TB approach can apply for other binary magnetic transition-metal alloys. It is expected to be particularly promissing if the size difference between the alloying elements is rather small and the electronic properties prevail.

\end{abstract}

\maketitle

\section{Introduction}
\label{intro}

Iron-Chromium systems have triggered extensive research efforts during the last few decades. On one side, it is due to their complex magneto-structural interplay, including the emergence of non-collinear magnetic configurations in the vicinity of structural defects and chemical heterogeneities \cite{Soulairol2011,Lavrentiev2011}. Magnetic interactions and frustrations are also shown to dictate thermodynamic properties such as the well-known atypical mixing-enthalpy behaviour of the Fe-Cr alloy \cite{Mirebeau1984,Mirebeau2010,Klaver2006,Levesque2011}. On the other side, these studies are motivated by the relevance of Fe-Cr systems for a large variety of technological applications. For instance, ferrito-martensitic steels 
with a high Cr content (around 10$\%$-Cr) show improved resistance to corrosion, irradiation and swelling. They are therefore promising materials for innovative nuclear devices. Also, FeCr multilayers were at the origin of the discovery of giant magnetoresistance\cite{Grunberg1986,Baibich1988}  which rapidly lead to tremendous application for electronic devices. 

Numerous atomic-scale studies based on density functional theory (DFT) have pointed out  a strong correlation between magnetic and energetic properties in the Fe-Cr alloys \cite{Ackland2006,Klaver2006,Ruban2008,Ropo2011,Olsson2007}. In the body-centered cubic (bcc) lattice, local magnetic moments on Fe atoms tend to be parallel (ferromagnetic), local moments on first nearest-neighbours ($1nn$) Cr atoms tend to be antiparallel (antiferromagnetic), and  within a simplified picture, moments of $1nn$ and $2nn$ Fe-Cr pairs prefer to be antiparallel \cite{Klaver2006,Lavrentiev2011}. Magnetic frustrations occur when these magnetic tendencies cannot be satisfied simultaneously. As a consequence, non-collinear magnetic configurations and/or spin-waves emerge in order to resolve partially the frustrations. This happens around the interfaces of Fe-Cr multilayers and of small clusters and precipitates in the binary alloy \cite{Soulairol2011,Lavrentiev2011,Fu2015}. Also, experiments and simulations indicated a mutual dependence of the microstructure and the global magnetization of the alloy \cite{Yamamoto1964,Fu2015}. In addition, the kinetics of phase decomposition in rather concentrated Fe-Cr alloys were shown to be very sensitive to the magnetic state of the system \cite{Martinez2012,Senninger2014}.        

Based on the above-mentioned evidences, an accurate description of the electronic structure and magnetism is essential for a reliable prediction of the thermodynamic, kinetic and defect and microstructural properties of the Fe-Cr alloys. Besides the first principles methods, semi-empirical interatomic potentials and models are often required for performing atomistic studies on systems containing defects (nano-clusters, dislocations, grain boundaries etc.), where large supercells, not reachable with DFT, should be adopted. In the case of Fe-Cr, empirical potentials based on the Embedded Atom model have been developed \cite{Olsson2005,Caro2005,DelRio2011}, where magnetic effects are taken into account only implicitly through the input parameters. It is not obvious that these potentials are able to predict the complex interplay between magnetism and energetic and structural properties of the defects. Tight binding (TB) models offer a promissing alternative, where the electronic structures and magnetism are explicitly considered. 
Previously, TB modelling of Fe/Cr interfaces were performed, addressing mainly the magnetic behaviour \cite{STOEFFLER1995,Martinez2006,Robles2003,Cornea2002}. More recently, a few tight binding models were developed paying special attention on the energetics and thermodynamics of Fe-Cr alloys 
\cite{Paxton2008,Nguyen-Manh2009,McEniry2011}. Attempting to go beyond, we present a new TB model, capable to predict both energetic and magnetic properties in the defect-free Fe-Cr alloys of different chemical compositions and ordering, and in the vicinity of surfaces, interfaces and nano-clusters. One specificity of this TB model is that no property from the binary system is explicitly included in the fitting data. In addition, the magnetism is treated beyond the usual collinear approximation, which is crucial for an accurate description of the FeCr system.                

The present paper is organized as follows: The TB formalism and the parameters fitting procedure are described in Sec. \ref{TB_model A} to \ref{TB_FeCr}. A comparison between the present and the previous TB models is given in Sec. \ref{TB_comparison}. In Sec.    \ref{properties}, we show the validity and transferability of the TB model by considering key properties of surfaces, interfaces, alloys of different compositions, the ordered B2 structure and small clusters, through a close comparison with the corresponding DFT results.
Finally, conclusions are given in Sec. \ref{conclusions}, and all the TB parameters are listed in the Appendix.

\section{A magnetic $spd$ tight binding model for alloys}
\label{TB_model}

\subsection{TB model for a single chemical-element system}
\label{TB_model A}

We have developed over the years an efficient scheme based on a tight-binding model  which we have extended to  spin-polarized systems\cite{Barreteau2000,Autes2006,Barreteau2016}. We will first recall the main ingredients of our model applied to single chemical element and then generalize it to metallic binary alloys.
The Hamiltonian is divided into three contributions:

\begin{equation}
\label{Eq:Hmag}
H=H^{\text{TB}}+V^{\text{LCN}}+V^{\text{Stoner}}
\end{equation}

$H^{\text{TB}}$ is the non magnetic TB Hamiltonian made of  diagonal elements $\epsilon_{i\lambda}= \langle i,\lambda | H | i, \lambda \rangle$ and hopping integrals $ \beta_{i\lambda, j \mu} =\langle i,\lambda | H | j, \mu \rangle$, where 
$|i,\lambda\rangle$ ($|j, \mu\rangle$) is the orbital $\lambda$ ($\mu$) on atomic site $i$ ($j$). The intra-atomic terms are written as a function of the local environment as in the work of Mehl and Papaconstantopoulos\cite{Mehl1996}.

\begin{equation}
\label{eq:intra}
\epsilon_{i\lambda}= a_{\lambda} + b_{\lambda} \rho_{i}^{1/3} + c_{\lambda} \rho_{i}^{2/3} +  d_{\lambda} \rho_{i}^{4/3}+ e_{\lambda} \rho_i^2
\end{equation}

$a_{\lambda}$, $b_{\lambda} $,  $c_{\lambda}$, $d_{\lambda} $ and  $e_{\lambda} $ are parameters to determine. $\rho_i$  is related to the atomic density around atom $i$ 

\begin{equation}
\label{eq:rhoLF}
\rho_{i}=\sum_{j \ne i} \exp[-\Lambda^2 r_{ij} ] F_c(r_{ij}).
\end{equation}

where the sum runs over the neighbouring sites $j$ surrounding atom $i$,
$\Lambda$ is  also a parameter and  $F_c(r)$ a cut-off function truncating interactions for distances larger than a given radius $R_c=16.5$Bohr, with a  Fermi-Dirac-like transition that brings  the potential smoothly to zero between $R'=14$Bohr and $R_c$.

The hopping integrals $ \beta_{i\lambda, j \mu}$ as well as the overlap integrals $ S_{i\lambda, j \mu} =\langle i,\lambda  | j, \mu \rangle$ are written in terms of 10 Slater Koster\cite{Slater1954} parameters $\beta_{\gamma}=$ $ss\sigma$, $sp\sigma$, $sd\sigma$, $pp\sigma$, $pp\pi$, $pd\sigma$, $pd\pi$, $dd\sigma$, $dd\pi$, $dd\delta$ which themselves are given an analytical form as a product of a decaying exponential and a polynome depending on  several parameters. 

\begin{equation}
\label{beta}
\beta_{\gamma}(r)=(p_{\gamma}+f_{\gamma}r+g_{\gamma}r^2)\exp[-h_{\gamma}^2 r]F_c(r)
\end{equation}

$V^{\text{LCN}}$ is the so-called "local charge neutrality" term that avoids  charge transfers by imposing a given electronic charge on each atom. The matrix elements of the corresponding potential  have the following form:

 \begin{equation}
 V^{\rm{LCN}}_{i \lambda \sigma, j \mu \sigma' }=\frac{1}{2}(U_i(N_i-N_i^0)+U_j(N_j-N_j^0))S_{i \lambda j \mu } \delta_{\sigma,\sigma'}
 \label{LCN}
\end{equation}

Where $N_i$ ($N_j$) is the Mulliken charge of atom $i$($j$) and $N_i^0$ ( $N_j^0$)the charge that one wants to impose on site $i$($j$).
$U_i$ depends only on the nature of the chemical element occupying site $i$ and determines the "strength" of the neutrality condition. $V^{\text{LCN}}$ is diagonal in spin space and acts indiferrently on up and down spins.

Finally $V^{\text{Stoner}}$ is the Stoner Hamiltonian which is the simplest but physically sound way to introduce magnetism in a tight-binding scheme. Its action is to split up and down bands in the following way
 \begin{equation}
 V^{\rm{Stoner}}_{i \lambda \sigma,  j \mu \sigma' }= -\frac{ I_{i \lambda}}{2}( M_{i d } \sigma  \delta_{\sigma,\sigma'})  \delta_{i \lambda,j \mu} 
 \label{Stoner-col}
\end{equation}

where $\sigma=\pm1$ denotes the up and down spin. $ I_{i \lambda}$ is the so-called Stoner parameter acting on orbital $\lambda$ and site $i$, and $M_{i d }$ is the component of the spin magnetization of atom on site $i$ summed over the $d$ orbitals only. In transition metals $d$ orbitals are the one bearing the magnetism and its value is controlled by the amplitude of  $ I_{i d}$ (the exact value of $ I_{i s}$ and $ I_{i p}$ has a minor effect on the total magnetization but in practice we took $I_s=I_p=I_d/10$).  $V^{\text{Stoner}}$ is diagonal in the spin-space and produces a shift between up and down spins. 

The Stoner potential can straightforwardly be generalized to non-collinear magnetism where the magnetization at each site can take any direction and must be described by a three component vector $\bm{M}_{i d}$. The potential now acts on both components of the spin-orbitals and can be written as a $2\times 2$ matrix:

 \begin{equation}
\tilde{ V}^{\rm{Stoner}}_{i \lambda,  j \mu  }= -\frac{ I_{i \lambda}}{2} (\bm{M}_{i d} . \bm{\sigma}) \delta_{i \lambda,j \mu}
 \label{Stoner-ncol}
\end{equation}
$\bm{\sigma}$ is the vector built from the three Pauli matrices $(\sigma_x,\sigma_y,\sigma_z)$ and the tilde denotes a $2\times 2$ matrix acting on a two component spin-orbital.

The total energy of the system is written in accordance with the work of Mehl and Papaconstantopoulos\cite{Mehl1996} as the sum of the occupied one electron eigenvalues $\varepsilon_{\alpha}$. This band term should however be corrected by the so-called double couting terms arising from electron-electron interactions introduced by LCN correction and Stoner terms\cite{Barreteau2012}. The total energy is then written as,

\begin{equation}
E_{\rm{tot}}=\sum_{\alpha}  f_{\alpha} \varepsilon_{\alpha} -\frac{1}{2}\sum_i U_i[N_i^2-(N_i^0)^2]+\frac{1}{4}\sum_{i,\lambda} I_{i \lambda}\bm{M}_{i \lambda}.\bm{M}_{i d}
\end{equation}

$f_{\alpha}$ being the occupation of state $\alpha$. The first term of the right handside expression is the so-called band energy of the  magnetic Hamiltonian given by Eq. \ref{Eq:Hmag} and the two other terms accounts for the double counting corrections arising from the local charge neutrality and Stoner potential.

Note that due to the electron-electron interaction  the Hamiltonian depends on the local charges and magnetic moments and thus the diagonalization of the Hamiltonian should be carried out self-consistently until the convergence criterium on the charge (and energy)  is achieved. 

It is worth mentionning that in the limit of large Coulomb interactions the term $U_i(N_i-N_i^0)$ converges towards a finite value $\delta V_i$ while $N_i$ approaches $N_i^0$. The double-counting correction term then takes 
the simple form $-\sum _i \delta V_i N_i^0$ valid in the limit of exact charge neutrality.

\subsection{Determination of TB parameters for single-element systems}
\label{TB_parameter}

The determination of the TB parameters is made in two steps. First, all the parameters of $H^{\rm{TB}}$ are obtained from a non-linear mean square fit to bulk nonmagnetic {\sl ab initio} band structures and total energy calculations for several lattice parameters and three different crystallographic structures (face centered cubic (fcc), body centered cubic (bcc)  and simple cubic (sc)) simultaneously.  An excellent agreement between TB and DFT results is obtained for both elements in the non-magnetic phase.

In a second step we determine the value of the Stoner parameter. This is done by a  trial and error approach where one tries to reproduce as precisely as possible the evolution of the magnetic moment $M$  of bulk material with the lattice parameter $a_{\rm{lat}}$ obtained from spin-polarized DFT calculations. Such calculations were performed on bulk bcc ferromagnetic (FM) iron and bulk bcc antiferromagnetic (AF) chromium. We found that $I_d^{\rm{Cr}}=0.82eV$ is a very good estimate for chromium while the case of iron is slightly more complex since it is difficult to reproduce the DFT results over the whole range of lattice parameters. Indeed  we found that for lattice parameters below 2.85{\AA}  the best value for the Stoner parameter is $I_{d}^{\rm{Fe}}=0.88$eV while for lattice parameters above 2.95{\AA} a larger Stoner parameter ($I_{d}^{\rm{Fe}}=0.95$eV) describes more accurately the $M(a_{\rm{lat}})$ curve. In addition the phase stability of iron is in much better agreement with ab initio data with $I_{d}^{\rm{Fe}}=0.95$eV than with $I_{d}^{\rm{Fe}}=0.88$eV, therefore all the calculations in this paper were performed with $I_{d}^{\rm{Fe}}=0.95$eV .

\subsection{phase stability of Fe and Cr}

\begin{figure}[!hbp]
\centering
\includegraphics[width=8cm]{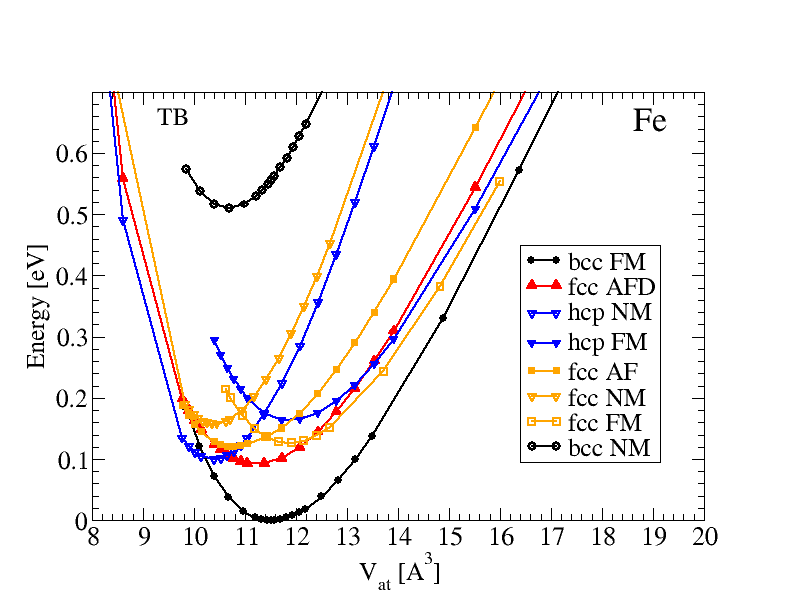} \includegraphics[width=8cm]{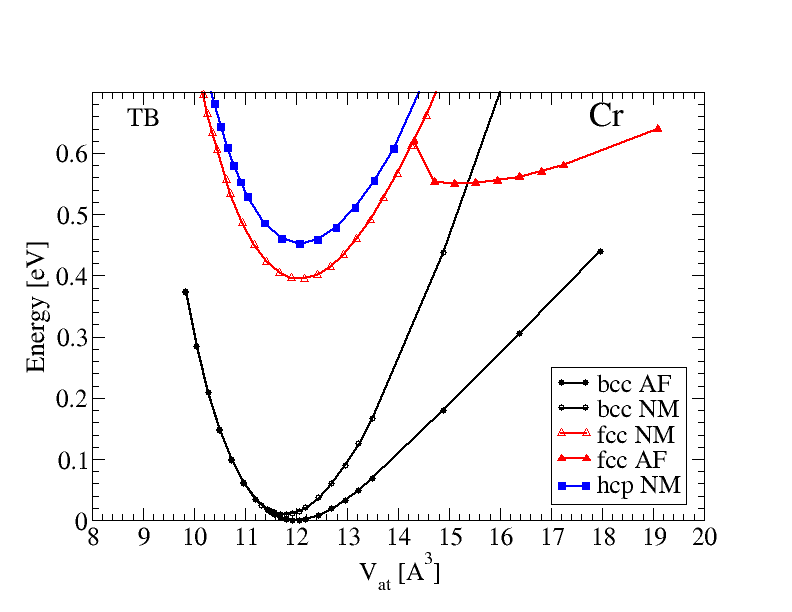} 
\caption{\label{fig:coheE}
Total energy (per atom) as a function of the atomic volume for various crystallographic structure (body centered cubic (bcc), face centered cubic (fcc) and hexagonal closed pack (hcp)), of Fe (left) and Cr (right). Several magnetic solutions are considered: non-magnetic (NM), ferromagnetic, simple layer antiferromagnetic (AF), double layer antiferrromagnetic (AFD). }
\end{figure}

Since our aim is to model Fe-Cr alloy over the whole concentration range it is essential to correctly reproduce the phase stability of both pure elements. This is particularly challenging for Fe since it  is known that even within DFT the choice of the functional can be crucial to accurately reproduce its phase stability. For instance it is well known that within local spin density approximation (LSDA) the non-magnetic hexagonal closed pack (hcp-NM) is found to be the most stable structure. It is only by using the generalized gradient approximation (GGA) that the most stable ferromagnetic body centered (bcc-FM)  is recovered. This is why we have fitted the TB parameters on GGA, DFT data.  
The relative stability of magneto-structural phases can be determined from energy versus atomic volume curves as plotted in Fig. \ref{fig:coheE} for Fe and Cr. The results are in surprisingly good agreements with DFT calculations of reference \onlinecite{Soulairol2010}. In particular the sequence in energy of the various phases of iron is almost prefectly reproduced. 

\subsection{TB model for FeCr binary systems}
\label{TB_FeCr}

\begin{figure}[!hbp]
\centering
\includegraphics[width=12cm]{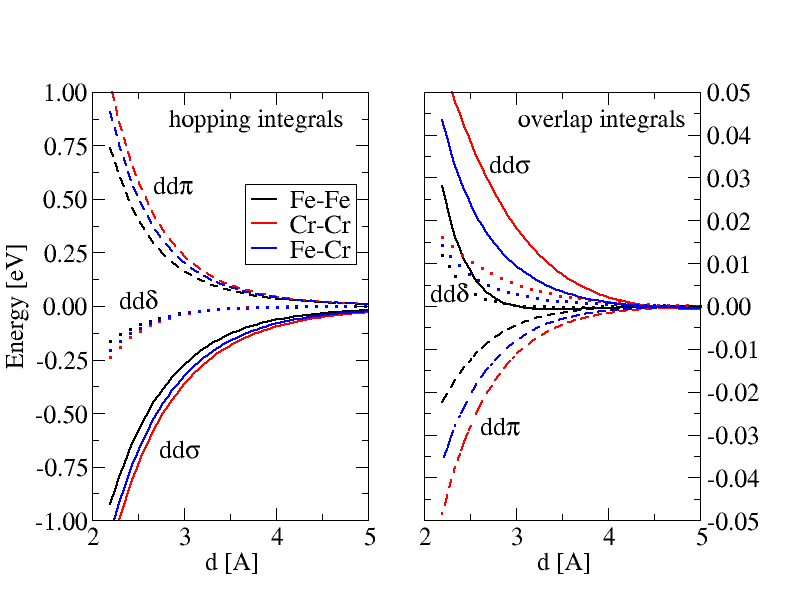} 
\caption{\label{fig:hopping-integrals}
Two-center Slater Koster $d$ hopping (left) and overlap (right) integrals as a function of the interatomic distance $r$ between two Fe-Fe, Cr-Cr and Fe-Cr atoms. The Fe-Cr hopping and overlap integrals are rescaled by a factor $1.023$ with respect to the arithmetic average.}
\end{figure}

If we consider now the Fe-Cr metallic alloy, the following procedure (which can be applied to any other transition metal alloy) has been carried out. A fit for both chemical element is performed separately with the same value of $\Lambda$. Then the intra-atomic terms of the Hamiltonian for a given site $i$ will only depend on the nature of the chemical specie occupying site $i$ by the value of the coefficients $a_{\lambda}$, $b_{\lambda} $,  $c_{\lambda}$ and  $d_{\lambda} $  for the corresponding atom. The hopping and overlap integrals between two identical atoms (Fe-Fe or Cr-Cr) are the same as the one abtained for the pure elements while the  hetero-nuclear value (Fe-Cr) is taken as the arithmetic average multiplied by a small (yet important) rescaling factor $\eta_{\text{Fe-Cr}}$

\begin{equation}
\label{Eq:beta-alloy}
\beta_{\gamma}^{\text{Fe-Cr}}(R) =\frac{\beta^{\text{Fe}}(R)+\beta^{\text{Cr}}(R)}{2}   \eta_{\text{Fe-Cr}}
\end{equation}

The distance dependence of the Slater Koster $d$ hopping and overlap integrals is illustrated in Fig.\ref{fig:hopping-integrals}. As expected the hopping (and overlap) integrals of Cr are larger than the one of Fe. In addition they can be very well approximated by a single exponential decay but this is not the case  for the integrals involving $s$ and $p$ orbitals (not shown in Fig.\ref{fig:hopping-integrals}).
The effect of $\eta_{\text{Fe-Cr}}$ is minor on the electronic and magnetic properties of the Fe-Cr alloy (magnetic moments are hardly affected by $\eta_{\text{Fe-Cr}}$) but the energetic of the alloy depends crucially on its numerical value. We found that $\eta_{\text{Fe-Cr}}=1.023$  gives the best results.

The role of the LCN term is evidently crucial in binay alloys since since it controls the charge on each atom. If the value of the Coulomb interaction $U$ is taken large enough the LCN condition is almost exactly fulfilled and the charge of a given atom of the system is equal to the valence charge of the corresponding element. In all this work we took  $U=30$eV for both elements which is sufficient to keep charge transfers as small as possible.
The Stoner parameter is taken as in the pure elements $I_{i,d}=I_d^{\text{Fe}}=0.95${eV} if site $i$ is occupied by an iron atom and and $I_{i,d}=I_d^{\text{Cr}}=0.82${eV} if site $i$ is occupied by a chromium atom.

Finally let us insist on the relative simplicity of our TB model for bi-metallic systems since it does not recquire any fitting to {\sl ab-initio} data from the binary alloy. The local charge neutrality condition aligns the respective local density of states so that the mulliken charge of each atom is close to the valence charge of the corresponding chemical specie. The hopping and overlap integrals are obtained from the ones of the pure elements. The only slight adjustment is related to the scaling factor $ \eta_{\text{Fe-Cr}}$ which  has a crucial influence on the energetics of the alloy, but a very small influence on its electronic and magnetic structure (at least for the very modest value taken in the case of Fe-Cr: $1.023$). In particular it is necessary to reproduce accurately the negative enthalpy of mixing for low Cr concentration of the Fe-Cr alloy. Let us also stress the generality of our procedure which can basically be applied to any alloy.

\subsection{Comparison with existing tigh-binding models}
\label{TB_comparison}

 In the past, several TB  modelling for Fe-Cr were performed but they were addressing mainly magnetic properties, in particular the frustration effects and non-collinear configurations at interfaces \cite{STOEFFLER1995,Martinez2006,Robles2003,Cornea2002}.
More recently, we are aware of essentially three magnetic tigth-binding models to describe both energetic and magnetic properties of the binary FeCr alloy: two are based on a $d$-band model\cite{Nguyen-Manh2009,McEniry2011}  and one on a $spd$-band model \cite{Paxton2008}. The two $d$-band model are very similar apart from details like the dependence of hopping integrals with interatomic distance which is exponential in Ref. \onlinecite{McEniry2011} or a power-law in Ref. \onlinecite{Nguyen-Manh2009}. The repulsive potential is also different since in Ref. \onlinecite{McEniry2011} an embedding potential is added to take into account the contribution of $s$-orbitals (ignored in Ref. \onlinecite{Nguyen-Manh2009}). The advantage of pure $d$-band models is evidently their simplicity and also their numerical efficiency. However one crucial parameter is the number of $d$ electrons  $N_d$ that should be defined  and when one is dealing with magnetic systems the choice of $N_d$ should be done concomitantly with the Stoner parameter $I_d$. There is evidently a rather large margin of choice since the two "parameters" are intimately connected. For example in Ref. \onlinecite{Nguyen-Manh2009}) they take $N_d=4.4$ and $I_d=0.7$eV for Cr while in Ref. \onlinecite{McEniry2011} it is $N_d=5.4$ and $I_d=0.54$eV. We believe that such a large variation of important parameters will necessarily lead to rather diffferent  physical behaviours in certain conditions, meaning that their transferabilty needs to be checked very carefully. In addition the electronic density of states are less faithfully reproduced than in a $spd$ model and several important features are often lacking. 

The $spd$-band model of Ref. \onlinecite{Paxton2008} is much closer to our model since not only it includes explicitely all the valence electrons but also takes into account overlap integrals.
The two $spd$ models essentially differ by two aspects: the distance dependence of their hopping and overlap integrals is simpler (exponential) than ours and the total energy is written as a sum of band and repuslive energy while we have adopted the procedure proposed by Mehl and Papaconstantopoulos\cite{Mehl1996} where the total energy (of a non-magnetic system) is written as the band energy only but on-site levels are varying with the local environment via Eq. \ref{eq:intra}. 

The specificity of our model is that we have fitted the hopping and overlaps integrals parameters to describe as closely as possible the band-structure and total energies of pure elements obtained from {\sl ab-initio} calculations on several crystallographic structures and over a large range of interatomic distances. We found this procedure important to reproduce accurately the complex intertwined magnetic and structural properties of Fe and Cr. For example our model perfectly reproduces the complex phase stability diagram in iron (see Fig. \ref{fig:coheE}) which is not possible with the simpler model of Ref.  \onlinecite{Paxton2008} and consequently it is not possible to study the mixing energies. In fact in the work of Paxton {\sl et al} the authors essentially focus on the magnetic contribution (Stoner like) to the total energy but not on the chemical contribution.

Our TB scheme has also been tested extensively to calculate magneto-crystalline anisotropy (therefore including spin-orbit coupling) in iron and cobalt with excelllent quantitative agreement with {\sl ab-initio} methods\cite{Li2013,Li2014}. Concerning the Fe-Cr alloy we have adopted a simple and straight forward procedure which proved to be very efficient and accurate.

\section{Model validation and results}
\label{properties}

Before discussing our results let us mention that in all our calculations we have only considered the standard antiferromagnetic (AF) configuration of bcc chromium and ignored any effect due to the spin-density wave (SDW) which is the true ground state of the material. 
We believe that the neglect of the SDW order has a modest influence on the following results since SDW and AF are very close energetically\cite{Soulairol2010} and it is known experimentally that the SDW phase disapears (in favour of the AF)  above a few percent of Fe incorporated in Cr. In addition structural relaxations are ignored. 

In the following, we compare systematically the surface energies of pure systems and various properties of the Fe-Cr alloys resulting from this TB model and predicted by our previous DFT studies \cite{Soulairol2011,Levesque2011,Fu2015}. The DFT calculations were calculated using the Siesta code \cite{siesta} within GGA in the Perdew-Burke-Ernzerhof (PBE) form. Core electrons are replaced by nonlocal norm-conserving pseudopotentials while valence
electrons are described by linear combination of numerical pseudo-atomic orbitals 
(LCAO). Either collinear or non-collinear treatment of magnetism has been adopted. A detailed description of the DFT simulation setup can be found in Refs. \onlinecite{Soulairol2011,Levesque2011,Fu2015}. 

\subsection{Surfaces}

\begin{table}[!hbp]
\begin{tabular}{|c|cc|cc|}
\hline
 Surface   & eV/atom & J/m$^2$   & eV/atom & J/m$^2$ \\
\hline
method                & \multicolumn{2}{c|}{TB}  & \multicolumn{2}{c|}{DFT-Siesta}\\
Cr(001)  & 1.606  &  3.091  &1.748  &3.380  \\
Fe(001) & 1.229    &  2.433  &1.474  &2.87   \\ 
Cr(110)  & 1.157   &  3.150  &1.258  &3.44   \\
Fe(110) & 0.750   & 2.100  &0.987  &2.70    \\
\hline
\end{tabular}
\caption{\label{Tab:surf}Surface energy per surface atom and per surface area for the $(001)$ and $(110)$ crystallographic orientations of bcc iron and chromium. Note that the surface energy (per surface area) of the $(001)$ orientation is lower than the one of the $(110)$ orientation in the case of chromium. This is attributed to a  very large enhancement of the magnetization on the outermost layer of bcc Cr$(001)$.  }
\end{table}

\begin{figure}[!hbp]
\centering
\includegraphics[width=10cm]{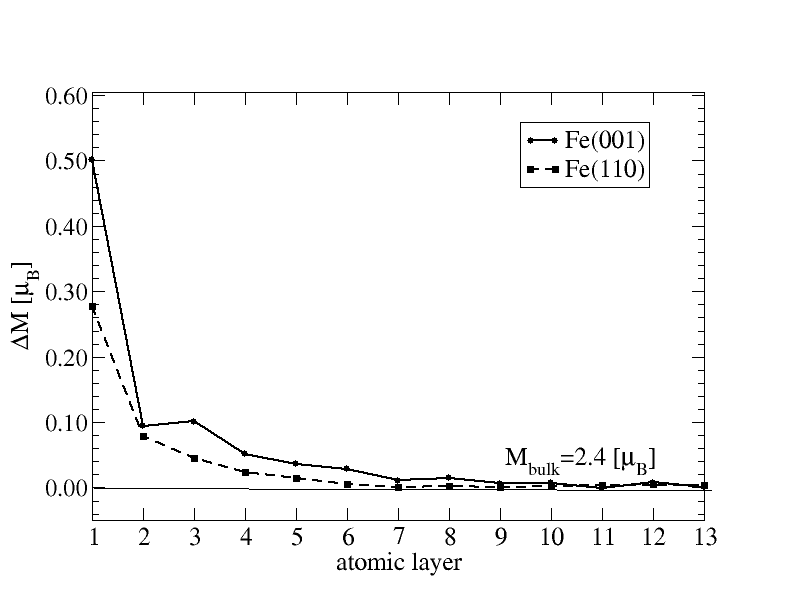} 
\caption{\label{fig:magFe}
Variation of the excess spin magnetization per atom (with respect to the bulk value)  decomposed on successive atomic layers for the $(001)$ and $(110)$ surfaces of Fe.}
\end{figure}

\begin{figure}[!hbp]
\centering
\includegraphics[width=10cm]{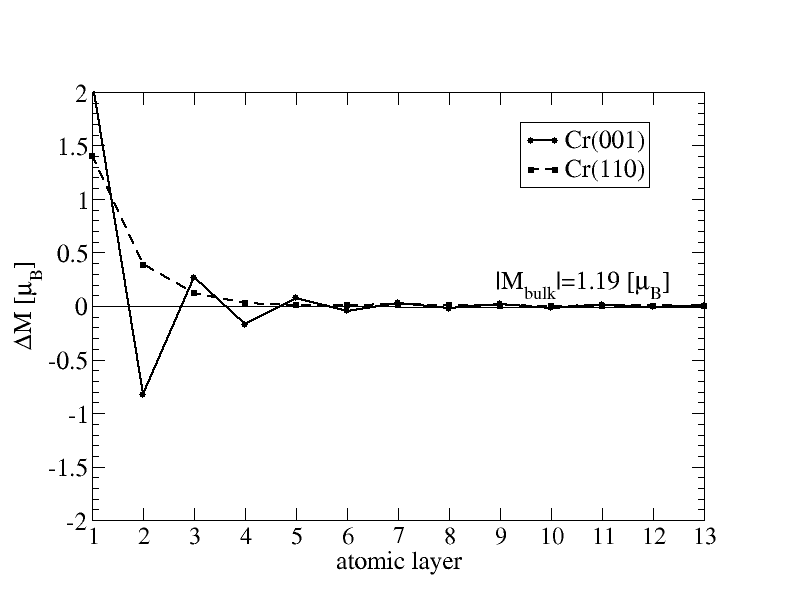} 
\caption{\label{fig:magCr}
Variation of the excess spin magnetization per atom (with respect to the bulk value)  decomposed on successive atomic layers for the $(001)$ and $(110)$ surfaces of Cr. Note the particularly strong enhancement of magnetization at the outermost layer of Cr$(001)$.}
\end{figure}

Before discussing the case of binary systems let  us first consider  the two lowest-index $(001)$ and $(110)$ surfaces of the pure elements. The surfaces are modelled by slabs of 27 atomic layers. Each atomic layer contains one atom per unit cell in the case of the $(001)$ orientation and two atoms  in the case of $(110)$. Therefore the slab of $(110)$ orientation contains twice more atoms $N^{\text{at}}$ than the one of $(001)$ orientation. In the case of iron the two atoms are equivalent and bear the same spin moment while for chromium they have opposite magnetic moments.
The lattice parameter is $a_{\text{bcc}}^{\text{Fe}}=2.845$ {\AA} for Fe and  $a_{\text{bcc}}^{\text{Cr}}=2.885 {\AA}$ for Cr and structural relaxation are ignored. 
The surface energies per surface area are calculated by the usual formula:

\begin{equation}
E^{\text{surf}}=\frac{1}{2A}\bigg[E^{\text{tot}}(\text{slab},N^{\text{at}})-N^{\text{at}} E^{\text{tot}}(\text{bulk}) \bigg]
\end{equation}
where $A$ is the  area of the surface unit-cell. $ E^{\text{tot}}(\text{slab},N^{\text{at}})$ and $E^{\text{tot}}(\text{bulk})$ denote the total energy of the slab (containing $N^{\text{at}}$ atoms) and of the bulk respectively. The numerical values are presented in Table \ref{Tab:surf}.

The magnetization is usually enhanced at surfaces as illustrated by Figs. \ref{fig:magFe} and  \ref{fig:magCr} showing the evolution of the spin moment as penetrating into the bulk of the material from its surface. This magnetization enhancement has consequences on their energetics  and can even modify the general trend for the surface energies. Indeed the surface energies of non-magnetic transition metals folllow the  rule of thumb based on the number of broken bonds that the densest surfaces have the lowest surface energies. However the surface magnetization follows an opposite rule that favors less dense surfaces since the more neighbours are lost at the surface the more the magnetization is increased with respect to the bulk. As a consequence,  less dense surfaces lower their energies by increasing their magnetization.  In the case of chromium $(001)$ the strong enhancement of the  surface magnetization $\Delta |M_{\text{surf}}|=|M_{\text{surf}}|-|M_{\text{bulk}}|=2\mu_B$ is strongly stabilizing this surface which energy (per surface area) is lower than the $(110)$ surface energy. The surface energies and in particular the energy difference between these two crystallographic orientations for both Fe and Cr are in good agreement with DFT results (Table \ref{Tab:surf}).

\subsection{Fe/Cr interfaces}

\begin{table}[!hbp]
\begin{tabular}{|c|cc|cc|}
\hline
 Surface   & eV/atom & J/m$^2$   & eV/atom & J/m$^2$ \\
\hline
method                & \multicolumn{2}{c|}{TB}  & \multicolumn{2}{c|}{DFT-Siesta}\\
(001) collinear         & 0.058  &  0.114  &0.062 & 0.120\\
(110) collinear        &  0.065   &  0.180   &0.073 &0.200 \\
(110) non collinear & 0.055   & 0.150   &0.069 &0.190 \\
\hline
\end{tabular}
\caption{\label{Tab:interface}Interface energy per interface atom and per interface area between Fe and Cr for the $(001)$ and $(110)$ crystallographic orientations.   Collinear and non-collinear magnetic structures are presented. Note that in the case of the $(001)$ interface all non-collinear initial configurations converge towards the most stable collinear configuration.}
\end{table}

\begin{figure}[!hbp]
\centering
\includegraphics[width=10cm]{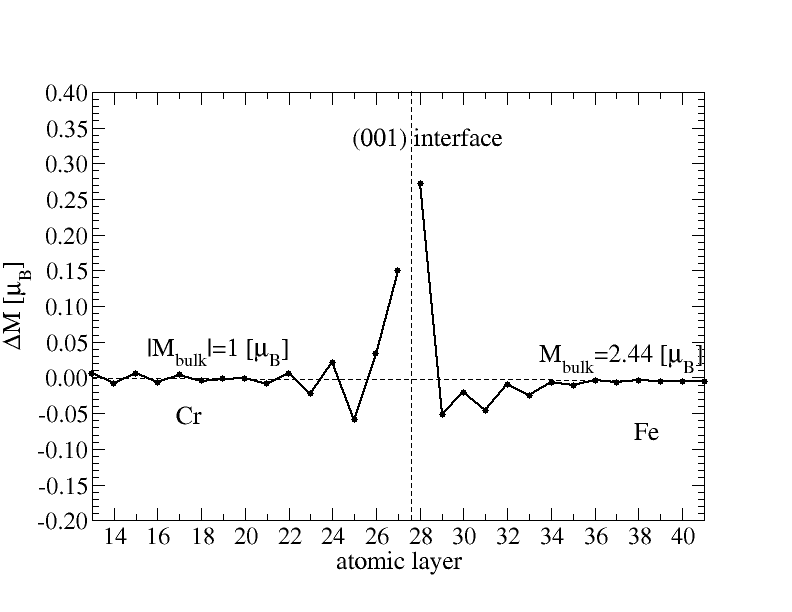}
\caption{\label{fig:magFeCr001}
Excess  local moments with respect to the corresponding bulk value ($M^{\text{bulk,Cr}}=\pm1 \mu_B$ and $M^{\text{bulk,Fe}}=2.44 \mu_B$) across the FeCr $(001)$ interface. }
\end{figure}

\begin{figure}[!hbp]
\centering
\includegraphics[angle=0, width=8cm]{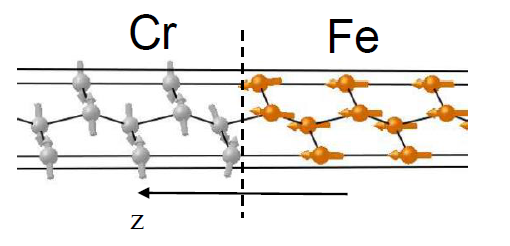}  
\includegraphics[width=10cm]{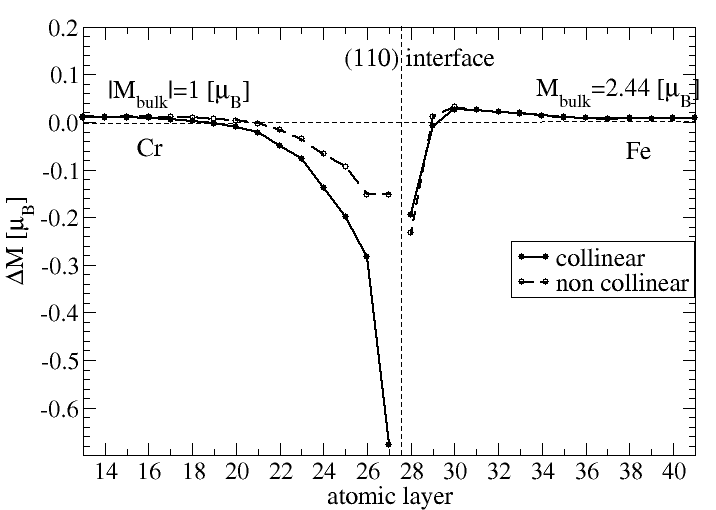}
\caption{\label{fig:magFeCr110}
Top: non-collinear magnetic configuration in the vicinity of the FeCr $(110)$ interface. Bottom: Excess  local moments with respect to the corresponding bulk value ($M^{\text{bulk,Cr}}=\pm1 \mu_B$ and $M^{\text{bulk,Fe}}=2.44 \mu_B$ ) across the FeCr $(110)$ interface for a collinear and and non-collinear magnetic solution. We note a rather modest canting of the magnetic moments of chromium and iron in the vicinity of the interface.}
\end{figure}

Let us now consider an interface between iron and chromium and investigate the role of magnetism on the energetics. In that respect  the $(001)$ and $(110)$ interfaces are expected to behave very differently since a strong frustration  is present at the $(110)$ interface due to the impossibility to fulfil the first neighbour antiferromagnetic coupling between Fe and Cr while at the $(001)$ interface such frustation does not exist (at least for the first neighbour interactions).  The systems are modelled by sticking together 27 layers of Fe and 27 layers of Cr. The lattice parameter is taken as the average value $a^{\text{int}}= \frac{1}{2}(a_{\text{bcc}}^{\text{Fe}}+a_{\text{bcc}}^{\text{Cr}})=2.865${\AA} and structural relaxations are ignored. The unit cell therefore contains 54 atoms in the case of the $(001)$ interface and twice more for the $(110)$ interface.
Collinear and non-collinear magnetic configurations are considered. For  non-collinear structures the starting magnetization is essential. We chose the initial magnetic moment of iron and chromium atoms  to be perpendicular and let  the system evolve until convergence was achieved. In the case of the $(001)$ interface the final configuration is the collinear one while for the $(110)$ a non-collinear magnetic solution do exist for which iron and chromium atom away from the interface have perpendicular magnetization while a small canting of the spin moments is observed in the vicinity of the interface. 

 The formation energy (per unit interface area)  of a given interface is then obtained from the formula:

 \begin{equation}
E^{\text{int}}=\frac{1}{2A}\bigg[E^{\text{tot}}(\text{int}, N^{\text{Fe}}/N^{\text{Cr}}  )-N^{\text{Fe}} E^{\text{Fe, tot}}(\text{bulk})- N^{\text{Cr}} E^{\text{Cr, tot}}(\text{bulk})\bigg]
\end{equation}

where $[E^{\text{tot}}(\text{int}, N^{\text{Fe}}/N^{\text{Cr}}  )$ is the total energy of the unit cell containing $N^{\text{Fe}}$ iron atoms and $N^{\text{Cr}}$ chromium atoms. $E^{\text{Fe, tot}}(\text{bulk})$ and$ E^{\text{Cr, tot}}(\text{bulk})$ are the respective bulk energy (per atom) of iron and chromium. The factor $2A$  accounts for the presence of two identical interfaces per unit cell.  The results are summarized in  Tab. \ref{Tab:interface}.

The obtained non-collinear ground state structure for the $(110)$ interface as well as the various interfaces energies from TB are in excellent agreement with DFT data. The lowest formation energy is obtained for the $(001)$ interface which can  be attributed to two concomitant mechanisms: i)  a stabilization of the $(001)$ interface due to an enhancement of the magnetization at the interface and ii) a strong frustration effect in play at the $(110)$ interface that decreases the amplitude of the magnetization at the interface and consequently penalizes the energetics of this interface. This frustration can be partly released by the development of a non-collinear magnetic configuration in the vicinity of the interface.

\subsection{The B2 phase} 

\begin{figure}[!hbp]
\centering
\includegraphics[width=5cm]{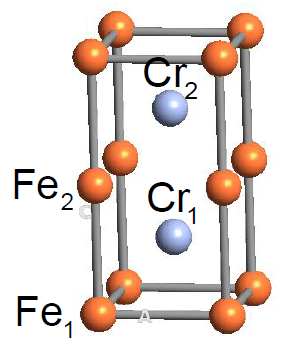}\includegraphics[width=10cm]{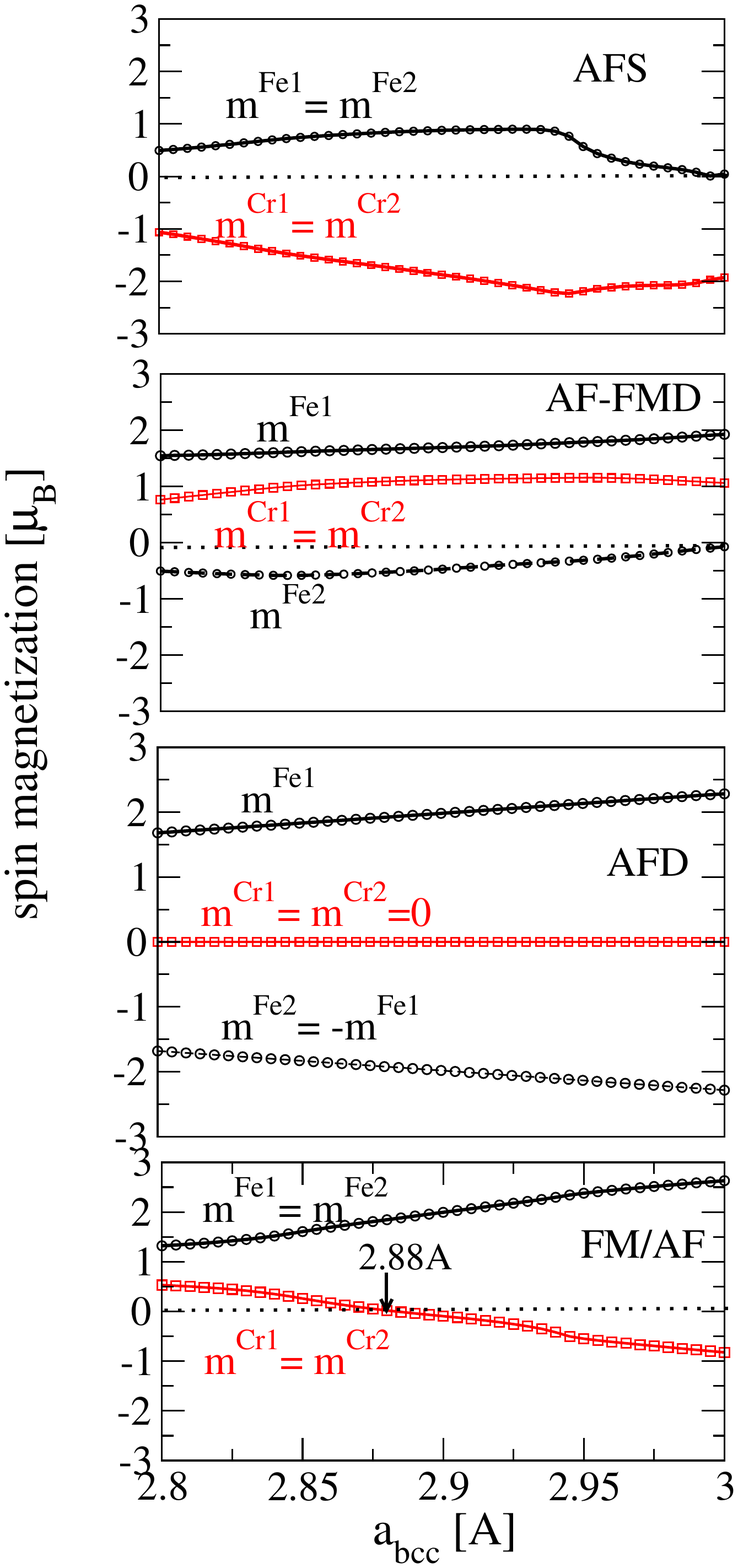} 
\caption{\label{fig:mag-vs-alat}
Left: Four-atom magnetic unit-cell used for our TB calculations of the FeCr B2 system. Right: Local Fe and Cr moments versus the bcc lattice parameter $a_{\text{bcc}}$ for the various magnetic phases. For the FM/AF phase the magnetization of chromium is switching sign at  2.88{\AA} }
\end{figure}

\begin{figure}[!hbp]
\centering
\includegraphics[width=8cm]{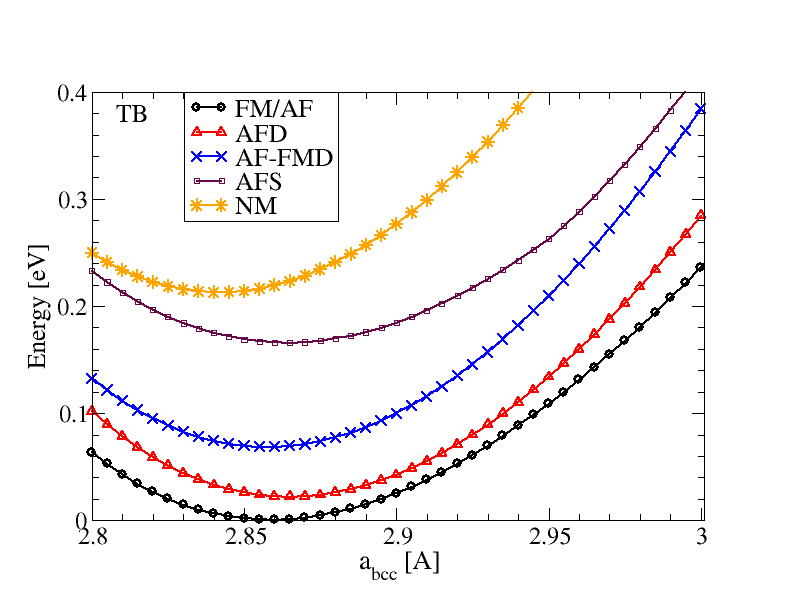} \includegraphics[width=8cm]{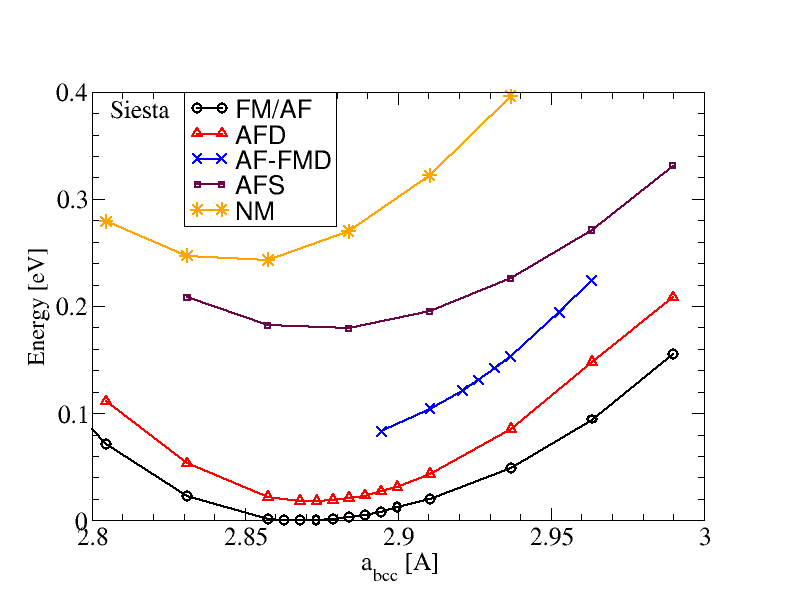}
\caption{\label{fig:E-vs-alat}
Calculated (TB: left. Siesta: Right) total energies (per formula unit) versus  the bcc lattice parameter $a_{\text{bcc}}$ for the various magnetic phases.  From Siesta calculations the AF-FMD structure cannot be obtained for lattice parameters smaller than 2.89{\AA}. For better comparison the minimum of the FM/AM curve has been set to zero in both calculations (TB and Siesta).}
\end{figure}

\begin{figure}[!hbp]
\centering
\includegraphics[width=8cm]{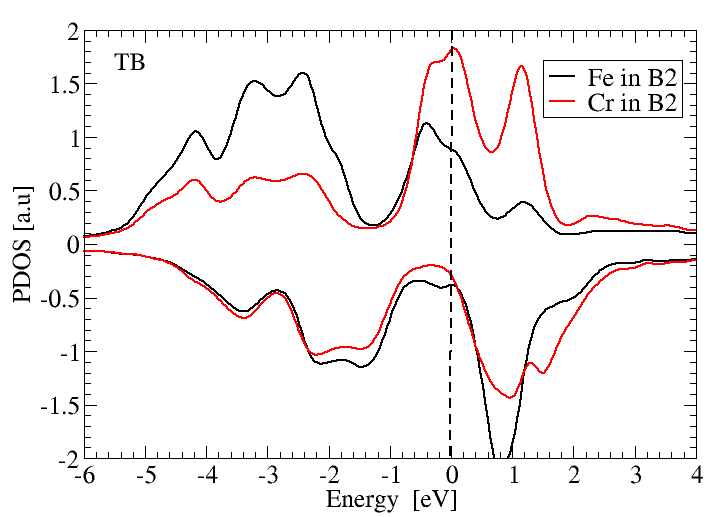} \includegraphics[width=8cm]{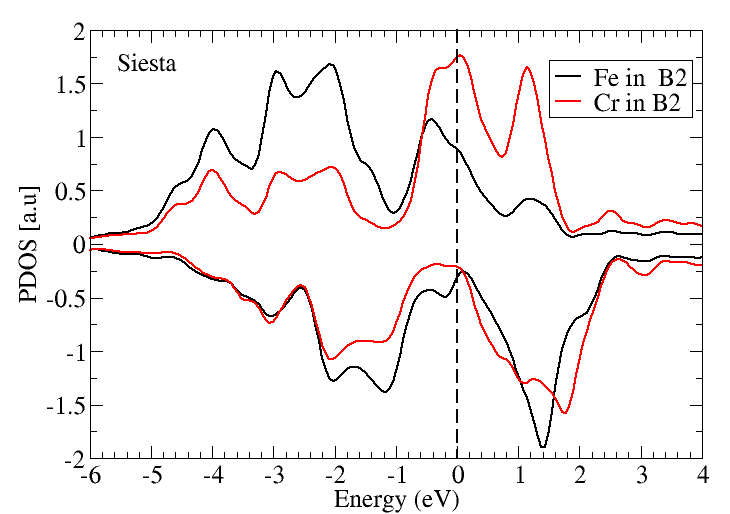} 
\caption{\label{fig:DOS-FeCrrB2}
Calculated (TB: left. Siesta: Right) spin-polarized density of states of states projected on the chromium (red line) and iron (black line) atomic orbitals for the Fe-Cr B2 FM/AF structure at the equilibrium lattice parameter $2.86${\AA}. Note that for this lattice constant Fe and Cr have both postitive magnetization. }
\end{figure}

The simplest  ordered crystallographic structure of the equiatomic Fe-Cr alloy is the B2 (or Cs-Cl) structure. The crystal system is simple cubic with two atoms per unit cell  based on the bcc lattice where one atomic specie occupies the corner of the cube and the other the center. This B2 phase has a very high formation energy in the case of Fe-Cr, but it is worth studying from the magnetic point of view. Indeed, although its  crystallography is very simple the magnetic structure of  Fe-Cr B2 is rather complex and several non-trivial solutions do exist in this phase. Inspired by the work of Qiu {\sl et. al.}\cite{Qiu1999} we have considered not only the B2 two-atom unit cell but also the four-atom magnetic unit-cell (see Fig. \ref{fig:mag-vs-alat}) built from two adjacent cubes in the $(001)$ direction. We have performed a careful investigation of the various magnetic structures by scanning many different initial magnetizations for the 4 (magnetically) inequivalent atoms in the  unit-cell. We have finally identified 4 different solutions (plus the non-magnetic one) in a given range of lattice parameters. Two of them (AFS and FM/AF) can be described by the elementary B2 unit-cell and the two others (AF-FMD and AFD) require the double four-atom unit-cell. Once these four solutions have been identified we have been  able to study their evolution with the lattice parameter in a range of lattice parameters around the equilibrium distance. In practice it was made possible to "follow" these solutions by performing a series of calcutations on a fine grid of lattice parameters starting from input charges and magnetization obtained from a previous solution.  In Fig. \ref{fig:mag-vs-alat} we show the evolution of the  local moments decomposed on the 4 different atomic sites of the double unit-cell for the four magnetic phases. And in Fig. \ref{fig:E-vs-alat} the corresponding total energy curves (per formula unit) are shown.  The lowest-energy solution is the so-called FM/AF  for which at the equilibrium distance (2.86{\AA}) both atoms have a positive magnetic moment below 2.88{\AA} while above this threshold the magnetization of Cr becomes negative. The closest solution in energy is the AFD solution for which the chromium atoms have a zero magnetic moment while iron atoms have large and opposite magnetizations. Just above in energy a rather unusual solution is obtained (AF-FMD) where both chromium atoms have the same positive magnetization while the two irons atoms have  moments of opposite signs:  a large and positive one and a small and negative one. The highest magnetic solution  in energy is  AFS where iron has a modest positive magnetization while chromium bears a large negative moment. Finally the non-magnetic solution is above all the magnetic ones showing that whatever the magnetic ordering the system always gain energy by developing some kind of magnetism. We have checked our TB results on the relative stability of the various magnetic phases of the B2 structure by performing DFT-Siesta calculations. The results of which are shown in Fig. \ref{fig:E-vs-alat}. The agreement between TB and DFT is once again excellent. It is worth mentioning that the phase stability obtained from our TB model differs significantly from the one of Qiu {\sl et. al.}\cite{Qiu1999} but we believe that it is due to the functional that they have used (LSDA) rather than a failure of our model. Indeed LSDA is known to overestimate bonding and consequently underestimate the latttice spacing which can strongly influence the phase stability of magnetic materials in particular in $3d$ transition metals. 

We have also calculated the projected  density of states of FeCr B2 in the FM/AF solution at the equilibrium lattice parameter $a=2.86${\AA} for which both magnetic moment of Fe and Cr are pointing in the same direction. Our TB results shown in Fig. \ref{fig:DOS-FeCrrB2} (left) are in very good agreement with the Siesta calculations (Fig. \ref{fig:DOS-FeCrrB2} right), proving the predictive character of our TB model not only for the energetics and magnetization but also for finer details of the electronic structure.

\subsection{Fe-Cr mixing enthalpies} 

\begin{figure}[!hbp]
\centering
\includegraphics[width=12cm]{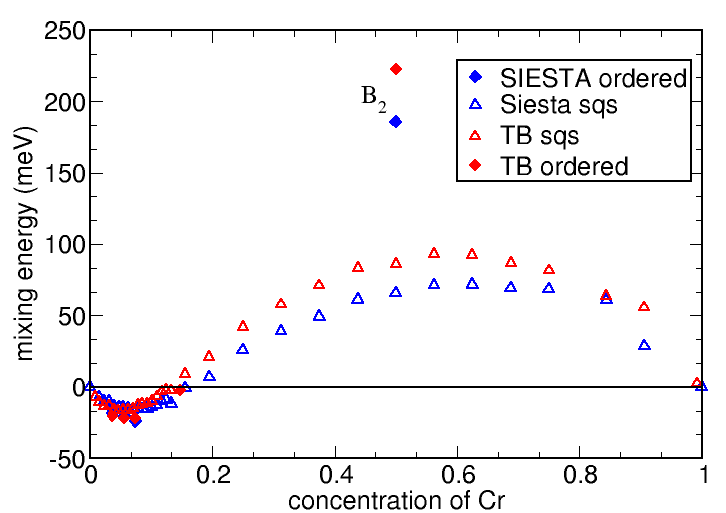} 
\caption{\label{fig:mixing-energy}
Enthalpy of mixing as a function of the concentration $c$ of Cr for the Fe$_{1-c}$Cr$_c$ alloy evaluated using our TB model compared to Siesta results. For each concentration $c$ we have used a sqs structure and only plotted the lowest energy solution when several magnetic configurations were found (esssentially in the high-concentration region). A few ordered structure were also calculated. }
\end{figure}

The enthalpy of mixing for the binary Fe-Cr alloy is known to present a specific negative feature for low concentration of chromium while it becomes positive for larger Cr content\cite{Olsson2003}. An accurate prediction of this behaviour is essential for studying any thermodynamic property of the alloy.  
We have calculated the  mixing enthalpy $\Delta H(c)$ of bulk Fe$_{1-c}$Cr$_c$ by considering a $4 \times 4 \times 4$ supercell of bcc lattice containing 128 atoms in total and varying the concentration $c$ of Cr from the lowest value ($1/128$) to the highest value ($127/128$). The lattice parameter of the alloy cell follows the Vegard's law $a(c)=c a_{\text{bcc}}^{\text{Cr}}+(1-c)a_{\text{bcc}}^{\text{Fe}}$. We have considered the special quasi-random structures (sqs)  which minimize the short-range order and are expected to be good representatives of solid solutions. $\Delta H(c)$ is evaluated by the standard formula:

\begin{equation}
\Delta H(c=\frac{m}{n+m})=\frac{E(Fe_nCr_m)-nE(Fe)-mE(Cr)}{n+m}
\end{equation}
Where $E(Fe_nCr_m)$ is the total energy of the supercell containing $n+m=128$ atoms and $E(Fe)$ , $E(Cr)$  are the equilibrium total energy per atom of bcc ferromagnetic Fe and bcc antiferromagnetic Cr. The results of our calculations are presented in Fig.\ref{fig:mixing-energy} and compared to Siesta calculations for the same set of structures. Our TB model reproduces very accurately the mixing-enthalpy curve over the whole range of concentration. In particular for the crossover between the region of negative and positive enthalpy the agreement is almost perfect.  The curve present a maximum for concentrations around 60\% of chromium. It is also important to note that in the region rich in chromium there often exist multiple magnetic solutions due to strong frustration (typically when two Fe are first neighbours) so that we had to test several initial magnetizations and only the lowest in energy was retained. In addition we have also considered the case of a few ordered structures essentially in the low Cr concentration region, where, as expected, the mixing enthalpy of these structures is always slightly more negative that the one of the sqs structure at the same concentration. The tendency is reversed for larger concentrations, where the mixing enthalpy is positive. This can be illustrated by the "pathological" case of the B2 structure which in a sense maximize the frustration and has the largest mixing-enthalpy.

\subsection{Small Fe (Cr) clusters embedded in a Cr (Fe) matrix.}

\begin{table}
\begin{tabular}{lcccc}
\hline
cluster                                        & $E_{mix}^{col}$  & $E_{mix}^{ncol}$ & $\Delta E_{tot}^{col-ncol}$  \\ [-2ex]
                                                  &  (meV/atom)          &  (meV/atom)         &   (meV/unit-cell)  \\
\hline
FeCr$_{127}$                                 &   2.4  &  -   &  0\\
Fe$_2$Cr$_{126}$                          &   5.6  &  4.8   & 95\\
Fe$_3$Cr$_{125}$   triangle           &  7.7    & 7.2 & 68\\ 
Fe$_4$Cr$_{124}$  tetrahedron     &  10.7    &  8.7 & 250\\
Fe$_4$Cr$_{124}$  square             &  14  &  13.8 & 47\\
\hline
CrFe$_{127}$                                  &    -7.2  &  -  & 0\\
Cr$_2$Fe$_{126}$                            &    -7.2  &  -  & 0\\
Cr$_3$Fe$_{125}$   triangle             &   -7.0    & -  & 0 \\ 
Cr$_4$Fe$_{124}$ tetrahedron       &   -5.8  & -   & 0\\
Cr$_4$Fe$_{124}$  square              &  -10.5  & -  & 0\\
\hline
\end{tabular}
\caption{\label{fig:cluster-energies}Mixing energies of various small Fe(Cr) clusters embedded in a Cr (Fe) matrix from TB collinear and non-collinear calculations. In the case of Cr clusters non collinear configurations do not exist while for Fe non-collinearity lowers the energy of the system. For a better comparison we have also listed the difference of energy between a collinear and a non-collinear configuration for a unit-cell of 128 atoms.  }
\end{table}

Finally, a good description of embedded clusters is necessary for studying for example properties of precipitates in concentrated Fe-Cr alloys, where there is a tendency for phase separation (positive mixing enthalpy) . Let us investigate two extreme cases: i) Small clusters of chromium in an iron FM bcc matrix and ii)  small clusters of iron in a chromium AF bcc matrix. Due to antagonists magnetic interactions we expect rather different behaviours in these two cases. Indeed the magnetic interaction between two iron atoms is FM while it is AF between two chromium atoms or between an iron and a chromium atom at near-neighboring positions. We have considered four different clusters: a dimer, a triangular trimer, a tetraheron and a square tetramer embedded in a $4\times 4$ supercell bcc latttice (the total number of atoms in the unit-cell being 128). Note that since the lattice is body centered the triangle and the tetrahedron are not regular since they connect either first or second neigbours. 
For each structure we have also investigated the possibility of occurence of collinear and non-collinear magnetic configurations. 
In Tables \ref{fig:cluster-energies} and \ref{fig:cluster-magnetization} we have summarized the energetics and the magnetization for the eight structures to which we have added the results for a single atom.

First let us note that no non-collinear configurations were found for these small Cr clusters. This is in agreement with previous DFT calculations  which showed that non-colinearity only appears for slightly larger clusters\cite{Fu2015}. In fact a chromium atom does favour the surrounding of iron atoms rather than chromium ones: For instance it is energetically more favorable (by about $0.4$eV) for two chromium atoms to be separated rather than first-neighbours. A large part of this Cr-Fe interaction is due to magnetism which is reflected by the strong enhancement of the magnetic moment on a single chromium atom ($-2\mu_B$) in an iron matrix compared to its bulk value ($\pm 1 \mu_B$) while in a dimer the magnetization of Cr drops to $-1.17\mu_B$. This is at the origin of the negative enthalpy of mixing for low Cr concentration indicating the tendency of Cr to make a solid solution. 

As soon as a chromium atom is connected to other chromium atoms the amplitude of its magnetic moment decreases rapidly. This is evidenced in the trimer where one Cr has two Cr atoms as first neigbours and the other two Cr have only one Cr as first neighbour (the other is a second neighbour). The magnetic moment of the single Cr connected to two other Cr is almost the same as in bulk Cr while the two other have a larger magnetization. The amplitude of magnetization of a Cr atom in tetrahedral geometry is almost the same as in the bulk while in the case of a square-shaped cluster the Cr atoms are second neigbours and bear a large magnetic moment as large as in the case of an isolated Cr. Interestingly for all these clusters the magnetic order between Cr atom is ferromagnetic proving that the Fe-Cr AF interaction is dominating the system. The AF magnetic order between Cr atoms will only be recovered for larger clusters when a sufficiently large number of Cr atoms have a bulk environment\cite{Fu2015}. 

In contrast for each iron cluster a lower energy non-collinear magnetic configuration does exist, in good agreement with DFT predictions\cite{Fu2015}. In addition, in most cases several collinear solutions were found but it is always the FM one (among Fe atoms) that is the  lowest in energy. This behaviour can be attributed  to a strong magnetic frustration which in fact do appear even for a single Fe atom in an AF Cr lattice since antiparalllel coupling cannot be fulfilled for both first and second Fe-Cr neighbours. Contrary to the case of Cr in Fe, the magnetic moment of the single Fe atom surrounded by only Cr atoms is strongly descreased compared to its bulk value. We found a magnetization of around 1$\mu_B$. In the case of the iron dimer in  the collinear configuration an asymmetric solution is found  where one atom has a zero magnetization while the other one is close to the iron bulk value. A symmetric solution is found in the non-collinear case where both iron atoms bear the same magnetic moment but canted with respect to one another (and to the Cr matrix). For the Fe trimer as in the case of Cr we found two Fe with large (negative) magnetization while the third Fe atom occupying the "up" sub-lattice of bcc-Cr has a lower (but still negative) magnetization. Similarly to the dimer in the non-collinear configuration all the iron atoms bear the same large magnetic moment  but canted with respect to one another. The iron tetrahedron is a very (magnetically) frustrated system as can be seen from the energy gain (250meV) by relaxing the collinear constraint to a non-collinear configuration. In contrast the square is a much less frustrated system but its mixing energy is higher since chemically iron prefers to form bonds with iron rather than with chromium.

\begin{table}
\centering
\includegraphics[height=3.0cm]{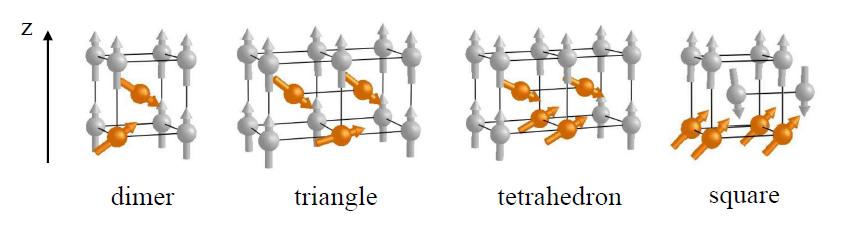}
\begin{tabular}{lcccc}
\hline
cluster                                                & $m_{\text{Fe(Cr)}}^{col}$ &$m_{\text{ Fe(Cr)}}^{ncol}$  & $\theta$      \\[-2ex]
                            & [$\mu_B$]  & [$\mu_B$] &  [degrees]\\
\hline
FeCr$_{127}$                                  &   -1.17  &  -  &  -\\
Fe$_2$Cr$_{126}$                          &   0/-2.28  &  1.89$\times$2   &  60/120\\
Fe$_3$Cr$_{125}$   triangle           &   -2.17$\times$2/-1.45    & 2$\times$3 &  130$\times 2$/80\\ 
Fe$_4$Cr$_{124}$  tetrahedron     &  -2.22$\times$2/-1.28$\times$2    &  2.08$\times$4 & 120$\times$2/60$\times$2 \\
Fe$_4$Cr$_{124}$  square             &   2.15$\times$4 &  2.07$\times$4 & 45$\times$4\\
\hline
CrFe$_{127}$                                     &  -2.3   &  -  & -\\
Cr$_2$Fe$_{126}$                            &  -1.19$\times$2    &  -  & -\\
Cr$_3$Fe$_{125}$   triangle             &   -1.73$\times$2/-1.26  & -  &  -\\ 
Cr$_4$Fe$_{124}$ tetrahedron       &   -1.25$\times$4  & -   &- \\
Cr$_4$Fe$_{124}$  square              &   -2$\times$4  & -  & -\\
\hline
\end{tabular}
\caption{\label{fig:cluster-magnetization}Magnetic moments for collinear and non-collinear magnetic coonfigurations of Fe(Cr) clusters embedded in a Cr(Fe) matrix. Whenever several atoms have the same magnetization we have indicated its multiplicity (for example $\times 4$ in the case of the four equivalent Fe atoms forming a square). At the top of the table we have shown the 4 different non-collinear configurations obtained for the iron clusters. }
\end{table}

\section{Conclusions}
\label{conclusions}

We have developed a new $spd$ TB model for the Fe-Cr system. The magnetism is treated within the Stoner formalism, beyond the usual collinear approcimation. A major advantage of this model consists in a rather simple fitting procedure. In particular, no specific properties of the binary system is explicitly required in the fitting database. Starting from the parameters of the pure systems, the hopping and the overlap integrals for the hetero-element (Fe-Cr) pairs are simply obtained by an arithmetic average multiplied by a unique rescaling factor. 

The resulting TB model is proved to be accurate and highly transferable for electronic, magnetic and energetic properties of a large variety of structural and chemical environments: surfaces, interfaces, embedded clusters, and the whole compositional range of the binary alloy. Note that none of these properties has been included in the fitting data.

It is worth mentioning that the present TB approach can apply for other binary magnetic transition-metal alloys. It is however particulary suitable for systems such as the Fe-Cr. Due to a very small size difference between Fe and Cr for instance in a bcc phase, several energetic properties of the alloy come to be driven by the electronic and magnetic interactions. The present TB model is also very promising if coupled with kinetic Monte Carlo simulations, for investigating finite-temperature magnetic and microstructural evolution in Fe-Cr alloys, where large-scale simulations are required.

       \acknowledgments
This work was performed using computer ressources from GENCI-DARI (Grant x2016096020), and IFERC-CSC within the SISteel project.  

\appendix*

\section{TB parameters}

In Tab. \ref{Tab:onsite} are listed the numerical values of the TB parameters to obtain the onsite elements of the Hamiltonian given by Eq. \ref{eq:intra} and \ref{eq:rhoLF}
in which the distances are expressed in Bohr and the energies in Rydbergs. The $\Lambda$ parameter (Eq.\ref{eq:rhoLF})  is taken equal to $1.3$ for both elements.

\begin{table}[!hbp]
\centering
\begin{tabular}{|c|c|ccccc|}
\hline
\hline
Element & Orbital &  \hspace{0.1cm} $a$ &   $b$ &   $c$ &  $d$  & $e$   \\
\hline
           &  $s$         &  0.0654     \hspace{0.1cm}   &  1.1144     \hspace{0.1cm}   &  11.4150    \hspace{0.1cm}   &   -469.0171     \hspace{0.1cm} &    7039.2378             \\
 Fe      &  $p$        &   0.3429      &  2.9992       &  -12.7329      &    157.7794     &    -880.7350             \\
           &  $d$        &   0.0744      &  -0.1788     &    1.6717       &       -2.1260      &      26.77154               \\
\hline
         &  $s$       &   0.0942      &    1.5564     &    5.1487         &     -267.1346   &      6295.1471          \\
 Cr    &  $p$       &   0.3343     &     4.3267     &   -21.3295      &      345.7256    &     -2234.9309           \\
        &  $d$       &   0.1135      &    -0.3014     &     4.1017      &       -21.2745   &    375.8615             \\
\hline
\end{tabular}
\caption{\label{Tab:onsite} Onsite TB parameters for Fe and Cr.}
\end{table}

In Tab. \ref{Tab:hopping} and \ref{Tab:overlap} are listed the numerical values of the TB parameters to obtain the Slater Koster hopping and overlap integrals of the Hamiltonian given by Eq. \ref{beta} in which the distances are expressed in Bohr and energies in Rydbergs. The  hetero-nuclear (Fe-Cr)  hopping (and overlap) integrals are  taken as the arithmetic average multiplied by the rescaling factor $\eta_{\text{Fe-Cr}}=1.023$  (See Eq. \ref{Eq:beta-alloy}).

\begin{table}[!hbp]
\centering
\begin{tabular}{|c|cccc||cccc|}
\hline
\hline
hopping &   \multicolumn{4}{c||}{Fe} & \multicolumn{4}{c|}{Cr} \\
\hline
                     &     $p$   & $f$   &  $g$ & $h$   &    $p$  & $f$  &  $g$ & $h$ \\
$ss\sigma$    &  0.0129  \hspace{0.1cm} & -0.7417   \hspace{0.1cm}   & 0.0392   \hspace{0.1cm} & 0.8020  \hspace{0.1cm}  &  0.3528   \hspace{0.1cm}   & -0.6590   \hspace{0.1cm}  & 0.0452  \hspace{0.1cm}   & 0.7572   \hspace{0.1cm}    \\
$sp\sigma$    &  -12.7214       & 3.7405   &  0.0304 &  0.9093 & -10.9485     &  2.8407 &  0.0836 & 0.9036 \\
$pp\sigma$   & -6.9952 & 2.4422 & -0.1802 & 0.7387  &  -8.3294    & 2.6866  &  -0.1647 & 0.7467 \\
$pp\pi$         & 148.7768 & -258.4013 & 0.000 & 4.4487  &  734.5209    & -98.8765  &  0.0000   &4.1281  \\
$sd\sigma$  &  2.2094 & -0.8765  & 0.0051 & 0.8878  &  3.6878    & -1.2032   & 0.028  & 0.8847 \\
$pd \sigma$ &  2.5908 & -1.0730 & 0.0589 & 0.8201 & 7.7230  &  -2.1013 &  -0.0054 & 0.9012 \\
$pd\pi$         &  -35.8525 & 11.8431 & -0.2706 & 1.1397 & -131.0844  &  39.6150 & -0.8188  & 1.2228 \\
$dd\sigma$ & -1.8022 & 0.3038 & -0.0164 & 0.7747 & -2.4171  &  0.3028 &  -0.0221 &  0.8357\\
$dd\pi$       &  6.6544  & -1.5783 & 0.1439 & 0.9635 & 5.6299 & -0.9789  & 0.0713  &  0.9314\\
$dd\delta$  & -0.0622 & -0.5314 & -0.0063 & 1.1286  & 14.0914  &  -6.7593 & -0.0344 &  1.2717\\
\hline
\end{tabular}
\caption{\label{Tab:hopping} Slater Koster hopping TB parameters for Fe and Cr.}
\end{table}

\begin{table}[!hbp]
\centering
\begin{tabular}{|c|cccc||cccc|}
\hline
\hline
overlap &   \multicolumn{4}{c||}{Fe} & \multicolumn{4}{c|}{Cr} \\
\hline
                     &     $p$   & $f$   &  $g$ & $h$   &    $p$  & $f$  &  $g$ & $h$ \\
$ss\sigma$    &  2.0429  \hspace{0.1cm} & -0.4161   \hspace{0.1cm}   & 0.2115   \hspace{0.1cm} & 0.8615  \hspace{0.1cm}  &  2.6878  \hspace{0.1cm}   &   -0.28736 \hspace{0.1cm}  & 0.1877  \hspace{0.1cm}   &  0.8581  \hspace{0.1cm}    \\
$sp\sigma$   &   0.6079 & -0.4843  & -0.0103  & 0.7465  & 2.4309   & -1.6073   & 0.0026  &   0.8052 \\
$pp\sigma$    &  3.8114  & -1.3166  &  -0.0014 & 0.7151  & 4.4633   &  -1.5723  & -0.0047  &  0.7554  \\
$pp\pi$          & -0.2540   & 1.9711  &  -0.0214 & 0.8594  &  -5.6357  &  2.7262  &  0.0072 &  0.8954  \\
$sd\sigma$    &  168.04884  & -25.9315  &  -2.4944 &  1.2560 &  3.74415  &  -0.7553  & 0.1202  & 0.9730   \\
$pd \sigma$   & 0.2049   & -0.2692  & 0.0348  &  0.6915 &  0.36665  & -0.0816   & 0.0099  & 0.6860   \\
$pd\pi$           &  -0.5420  &  0.0992 & -0.0046  & 0.4195  &  -0.6352  &  -0.2187  & -0.0012  &  0.8107  \\
$dd\sigma$    & 22.7769   & -1.2565  & -0.4900  & 1.1789  & -0.90857   & 0.8767   & -0.0911  &   0.8521 \\
$dd\pi$           &  3.6198  & -1.5098  & -0.4374  & 1.2132   & -2.0957   & 0.2115   &  -0.0017 &  0.8834  \\
$dd\delta$      & 10.2436  & -0.5319  & -0.1977  &  1.1980 & 0.2764   &  -0.0260  & -0.0001  &   0.7488 \\
\hline
\end{tabular}
\caption{\label{Tab:overlap} Slater Koster overlap TB parameters for Fe and Cr.}
\end{table}

\clearpage

\bibliographystyle{apsrev}
\bibliography{FeCr}

\clearpage


\end{document}